\documentclass[aps,prl,twocolumn,letterpaper,superscriptaddress,showpacs]{revtex4-2}
\usepackage{graphicx}
\usepackage{CJK}
\usepackage{amsfonts}
\usepackage{verbatim}
\usepackage{physics}
\usepackage[colorlinks,linkcolor=blue,anchorcolor=blue, citecolor=blue,urlcolor=blue,]{hyperref}
\usepackage{lineno} 
\usepackage{bm}
\usepackage{mathptmx}
\usepackage{amsmath}
\usepackage{textcomp}
\usepackage{xcolor}
\usepackage[normalem]{ulem}
\usepackage{gensymb}  
\makeatletter

\begin{document}
\begin{CJK*}{UTF8}{gbsn}

\title{Quantization concealment of angular momentum: dynamic origin of spins}

\author{Wenjie Long (\CJKfamily{gbsn}龙文婕)}
\affiliation{School of Physics and Astronomy, Shanghai Jiao Tong University, Shanghai 200240, China}
\author{Wei Ku ({\CJKfamily{bsmi}顧威})}
\altaffiliation{corresponding email: weiku@sjtu.edu.cn}
\affiliation{School of Physics and Astronomy, Shanghai Jiao Tong University, Shanghai 200240, China}
\affiliation{Key Laboratory of Artificial Structures and Quantum Control (Ministry of Education), Shanghai 200240, China}
\affiliation{Shanghai Branch, Hefei National Laboratory, Shanghai 201315, China}

\date{\today}

\begin{abstract}
Since their introduction, spins of elementary particles remain a phenomenological addition to modern quantum theories.
While relatively recent studies on Dirac and other linearized equations of motion have revealed the \textit{spatial current} nature of the spin magnetic moment (instead of the commonly pictured self-rotation), the dynamical origin of this circulating current remains unresolved.
Here, we expose a long-overlooked generic shortcoming of wave mechanics -- `quantization concealment of angular momentum' -- that undesirably reduces the canonical angular momentum of bound eigen-modes and renders incomplete the minimal coupling to magnetic fields.
Recovering the kinetic \textit{orbital} angular momentum resolves $1s$-orbital's counterintuitive ``zero'' angular momentum and reproduces all planar modes' anomalous Zeeman splitting.
Regarding spins as effective remedy to the concealment offers natural explanations to their main puzzles: 1) dynamical origin of spin magnetic moment and 2) frame invariance of spin angular momentum.
This implies \textit{absence} of spin magnetic moment for free particles, a strong prediction verifiable through Stern-Gerlach experiments on electrons.
\end{abstract}

\maketitle
\end{CJK*}

\textit{Introduction} --
Centuries after the development of quantum theories, spins of elementary particles remain their most mysterious properties.
Originally, spins were \textit{phenomenologically} introduced to explain the anomalous Zeeman effect\,\cite{UhlenbeckGoudsmit1926,THOMAS1926,KRONIG1926}, a frequency splitting due to magnetic moment's coupling to magnetic field, observed even in hydrogen's ``zero'' orbital angular momentum $1s$ orbital.
Subsequently, combined with the Pauli principle\,\cite{Pauli1925}, the inclusion of 1/2-spin also provides a group structure\,\cite{Pauli1927, Dirac1926,dirac1981principles} consistent with the periodic table.
Yet, despite spins' essential roles in quantum theories, to date its microscopic \textit{dynamic} origin remains unresolved.

Theoretically the most natural introduction of spins is through the extended Hilbert space in linearizing the equations of motion\,\cite{Dirac1928, levy1967nonrelativistic}.
However, unlike the Hilbert space corresponding to the spatial degrees of freedom, such extended space has no direct correspondence to any physical degrees of freedom.
Correspondingly, the ``spinor'' Hilbert space only \textit{mathematically} serves the linearized equations of motion, without physical association to any established dynamics before linearization.
Consequently, their common physical pictures are either pure mathematical or products of preferred imagination: self-rotation being the most popular one.

Interestingly, even within the framework of such linearized equations of motion, previous studies\,\cite{Ohanian,Gordon1928,belinfante1939,belinfante1940} have unambiguously revealed a \textit{spatial} natural of spin magnetic moment in association with a circulating current inside any wave packet (instead of particles' self-rotation).
Other than the obvious question on the dynamical \textit{origin} of such circulating current present even in a freely propagating wave packet, many unnatural properties of spins remain puzzling.
For example, the spin angular momenta is strangely \textit{inertial frame invariant}, contrary to regular angular momenta.
Yet another, independent of their size, $S$, in such framework spins' gyromagnetic ratio is always $g_s=S^{-1}$\,\cite{Hagen1970,belinfante1953,Holstein2006}, such that their magnetic moment, $\frac{e\hbar}{2m_e}$, is \textit{independent} of their size.
Given spins' essential rules in quantum theories, a physical resolution to these puzzles is conceptually of utmost importance.

Here, we expose a long-overlooked generic shortcoming of wave mechanics -- `quantization concealment of angular momentum' -- that reveals the dynamical origin of this spin-associated spatial current of bound eigen-modes and explains the frame invariance of spin angular momentum.
Specifically, each quantized dimension beyond 1D hides a $\frac{\hbar}{2}$ piece of the \textit{kinetic} angular momentum of bound eigen-modes.
We show that the properly recovered kinetic \textit{orbital} angular momentum naturally resolves $1s$-orbital's counterintuitive ``zero'' angular momentum and reproduces all planar modes' anomalous Zeeman splitting.
Viewing spins as effective remedy to the concealment offers natural explanations to their unresolved puzzling properties.
Our finding implies \textit{absence} of spin magnetic moment for free particles, a strong prediction verifiable through Stern-Gerlach experiments on electrons.

\textit{Spin-associated spatial current} --
Recall that a spin, $\mathbf{S}$, can be naturally introduced by extending the Hilbert space with a ``spinor'' of preferred size in linearizing the equations of motion\,\cite{Dirac1928, levy1967nonrelativistic}, even in \textit{non-relativistic} theories.
Importantly, the spin magnetic moment, for example $g_s\mathbf{S}\equiv\mathbf{\mu}_\mathrm{s}=-\frac{e\hbar}{2m_e}\sigma$ of the 1/2-spin of an electron of mass $m_e$ and charge $-e$, is produced \textit{not} through the commonly pictured self-rotation, but via an spin-associated \textit{spatial} current density,
\begin{equation}
\mathbf{j}_\mathrm{s}(\mathbf{r})=\frac{\hbar}{2m_e}\nabla\times(\psi^\dagger(\mathbf{r})\sigma\psi(\mathbf{r})),
\label{eq:NR_current}
\end{equation}
in addition to the standard orbital current encoded in the two-component wave function $\psi(\mathbf{r})$, as easily verified by its contribution to the magnetic energy,
\begin{equation}
U_s =-\int(-e)\mathbf{j}_\mathrm{s}\cdot \mathbf{A}~d^3r
=-\mathbf{B}\cdot\int\psi^\dagger(\frac{-e\hbar}{2m_e}\sigma)\psi~d^3r,
\label{eq:NR_moment}
\end{equation}
under a magnetic field, $\mathbf{B}$, or its corresponding potential, $\mathbf{A}$.
(Here $\mathbf{\sigma}$ denotes the vector of Pauli matrices.)
This \textit{spatial orbital} nature of the spin magnetic moment is, in fact, \textit{independent} of relativistic effects, as a similar spin-induced spatial current also results\,\cite{SM} from Gordon decomposition \,\cite{Gordon1928} of the Dirac equation.

Nonetheless, it remains unclear why such spatial angular momentum is inertial frame invariant\,\cite{Saradzhev2011,levy1967nonrelativistic}, contrary to regular angular momenta.
Similarly, it is rather curious that after all these years, there is yet a successful \textit{direct} confirmation for the existence of a free electron's spin magnetic moment through the Stern-Gerlach experiment.
Even with potential obscuration~\cite{pauli1932proceedings,pauli2012general,mott1929scattering} from the uncertain principles~\cite{pauli1932proceedings,pauli2012general}, there should still be a \textit{detectable} external magnetic field gradient-\textit{dependent} spectral spread~\cite{batelaan1997stern,karimi2012spin}.
Most essentially, the microscopic \textit{dynamical} origin of such an circulating momentum flow\,\cite{Bliokh2014,Mita2000} remains unknown to date.

\begin{figure}[t]
  \centering
  \vspace{-0.2cm}
  \includegraphics[width=\linewidth]{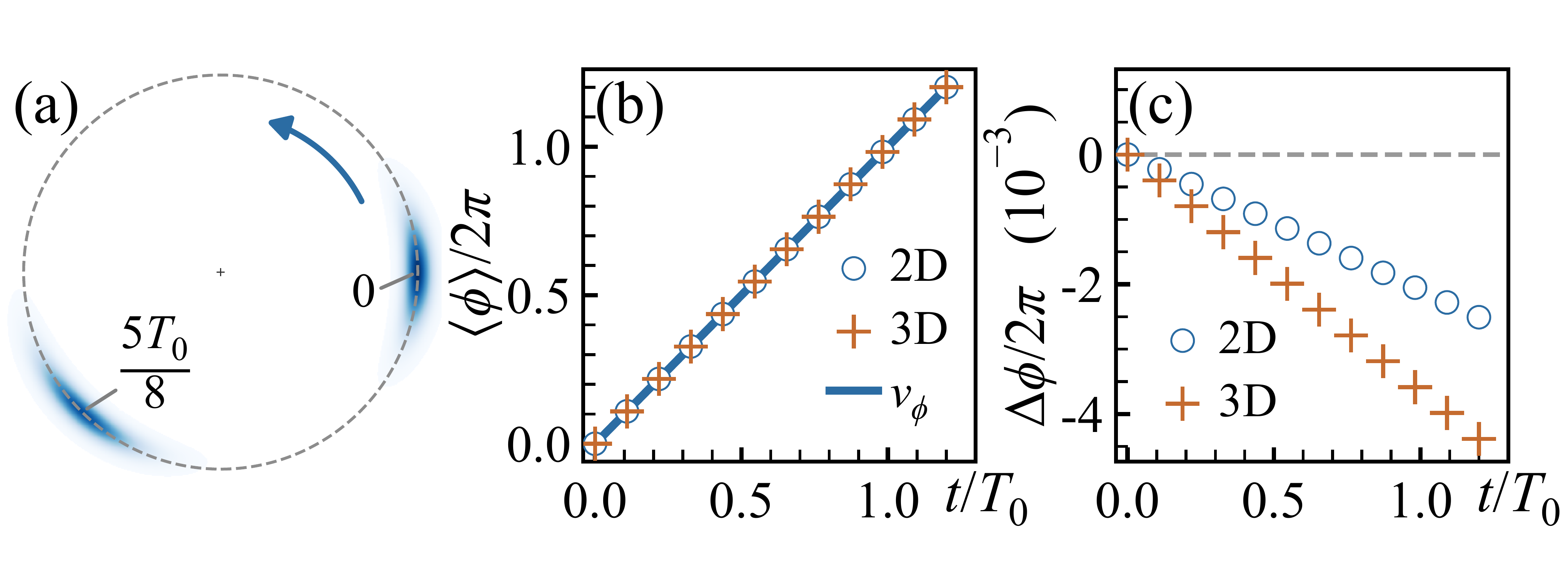}
  \vspace{-.8cm}
  \caption{\textit{Demonstration of quantization concealment} -- (a) Time evolution (in unit of the period, $T_0$) of circularly orbiting wave packets under 2D and 3D quantization, having distribution of width $\Delta =3$ centered at $n_0 = 320$.
  (b) $\langle \phi(t) \rangle$ in both cases evolve with a constant angular velocity nearly identical to $v_\phi$ in Eq.\,\ref{eq:group_angular_velocity} (or the Bohr model).
  (c) Angular deviations, $\Delta\phi\equiv \frac{\langle L\rangle}{(m_e r_0^2)}t-\langle \phi \rangle$, show that the angular velocity is indeed underestimated by the canonical $L$ in 2D quantization, and twice more in 3D quantization.}
  \label{fig:circular_2d_panels}
  \vspace{-0.4cm}
\end{figure}

\textit{Consistent 2D and 3D quantizations} --
As a representative demonstration, consider a central force system with an electron in vacuum (dielectric constant $\epsilon_0$) of mass $m_e$, charge $-e$, and momentum $\mathbf{p}$ at location $\mathbf{r}$, described by a Hamiltonian,
\begin{equation}
  H = \frac{p^2}{2m_e} - \frac{e^2}{4\pi\epsilon_0}\frac{1}{r}.
\label{eq:H_central_force}
\end{equation}
Upon quantizing all three spatial dimensions and utilizing the rotational symmetry, for \textit{bound} eigen-modes the problem is reduced to the eigen-equation for radial function, $u_{nl}(r)$,
\begin{equation}
\frac{\hbar^2}{2m_e} \left( -\partial_r^2 -\frac{2}{r}\partial_r + \frac{l(l+1)}{r^2} -\frac{2}{a_0}\frac{1}{r} \right) u_{nl}(r) = \hbar \omega_n u_{nl}(r),
\label{eq:3dradial}
\end{equation}
for any given angular quantum number $l$, with eigen-frequency, $\omega_n$, where $a_0\equiv\frac{4\pi\epsilon_0\hbar^2}{m_ee^2}$ denotes the Bohr radius.

According to the Heisenberg equation of motion, this system has a symmetry above and below the plane hosting $\mathbf{r}$ and $\mathbf{p}$, such that orbiting motion stays in that plane.
Therefore, one expects consistent description upon quantizing just the two dimensions of the planar motion, with the corresponding eigen-equation for the radial function, $v_{n|m|}(r)$,
\begin{equation}
\frac{\hbar^2}{2m_e} \left( -\partial_r^2 -\frac{1}{r}\partial_r+\frac{m^2}{r^2}-\frac{2}{a_0}\frac{1}{r} \right) v_{n|m|}(r) = \hbar \omega_n v_{n|m|}(r),
\label{eq:2dradial}
\end{equation}
for any given angular wavenumber $m$.
Indeed, one finds\,\cite{suslov2024sommerfeld} that setting $v_{nm}(r)$ as $r^{\frac{1}{2}}u_{n,l=|m|-\frac{1}{2}}(r)$ in Eq.\,\ref{eq:2dradial} exactly \textit{reproduces} Eq.\,\ref{eq:3dradial}.
That is, the 2D-quantized description \textit{perfectly} reproduce all the planar dynamics of the standard 3D quantization, including $\omega_n$ and radial probabilities, $\rho_{nl}(r)\equiv|u_{nl}(r)|^2 r^2=\left. |v_{n|m|}(r)|^2 r\right|_{|m|=l+\frac{1}{2}}$, as long as the angular wavenumber $m$ is quantized via $|m|=l+\frac{1}{2}$.
(Such a half-integer $m$ corresponds to a twist\,\cite{Byers1961} in the $\phi=2\pi$ to 0 boundary, and is easily ``gauge transformed''\,\cite{Akkermans1997} to a periodic $\ket{\tilde{\phi}}$ via $\ket{\tilde{\phi}}=e^{i\phi/2}\ket{\phi}$.)

To see the consistent dynamics of these two descriptions, consider the time evolution of a normalized circularly orbiting wave packet under 3D and 2D quantizations,
\begin{align}
  &\psi^\mathrm{(3D)}(r,\theta,\phi;t) \propto \sum_{n=1}^{\infty} \psi_n u_{nl}(r) \sin^l(\theta) e^{i(m\phi -\omega_nt)}\bigg|_{m=l=n-1} \nonumber \\
  &\psi^\mathrm{(2D)}(r,\phi;t) \propto \sum_{n=1}^\infty\psi_n v_{n|m|}(r)e^{i(m\phi-\omega_nt)}\Big|_{m=n-\frac{1}{2}}
\label{wave_packet}
\end{align}
in the weakly bound  ($n \gg 1$) regime via distribution, $  \psi_n=e^{-\frac{(n-n_0)^2}{4\Delta^2}}$,
of width $\Delta$ centered at $n_0$.
Given $\omega_n=-\frac{\hbar}{2m_ea_0^2} n^{-2}$ and the most probable orbital radius, $r_n= a_0n^2$, both wave packets have the same angular group velocity,
\begin{equation}
  v_\phi\equiv \frac{d\omega_n}{dm}\bigg|_{n_0}=\frac{\hbar n}{m_er_n^2}\bigg|_{n=n_0},
\label{eq:group_angular_velocity}
\end{equation}
in perfect agreement with (1D quantized) Bohr's model.
These results are easily confirmed numerically.
Figure.\,\ref{fig:circular_2d_panels}(b) shows the numerical results on time evolution of such wave packets with $n_0=320$ and $\Delta=3$.
Both wave packets indeed propagate in a circular orbit with average angular velocity, $\frac{d}{dt}\langle \phi \rangle$, nearly identical to $v_\phi$ in Eq.\,\ref{eq:group_angular_velocity}.

\textit{Quantization concealment of angular momentum} --
Notice that such consistency between 3D and 2D quantizations is accompanied by an overall $|m|-l=\frac{1}{2}$ reduction in the angular wavenumber under 3D quantization, or an $\hbar |m| -\hbar l =\frac{\hbar}{2}$ reduction (in each orbiting direction) in the canonical angular momentum, $L\equiv -i\hbar\partial_\phi$.
Obviously, given the identical $\omega_n$, $\rho_{nl}(r)$, and wave packet dynamics, such overall reduction must have \textit{no} physical implication and the shift-able \textit{canonical} angular momentum, $L$, must differ from the physical \textit{kinetic} angular momentum, $L^\mathrm{(K)}$.

\begin{table}
\centering
\begin{tabular}{l||c|c|c||c|c} \hline
& \multicolumn{3}{c||}{central force} & \multicolumn{2}{c}{cyclotron} \tabularnewline \hline
& Bohr (1DQ) & 2DQ & 3DQ & $~~~$ 1DQ $~~~$ & 2DQ / +$z$Q \tabularnewline \hline
$m$ & $n$ & $n\textcolor{red}{-\frac{1}{2}}$ & $n\textcolor{red}{-1}$ & $n$ & $n\textcolor{red}{-\frac{1}{2}}$ \tabularnewline \hline
$L$ & $\hbar n$ & $\hbar n\textcolor{red}{-\frac{\hbar}{2}}$ & $\hbar n\textcolor{red}{-\hbar}$ & $\hbar n$ & $\hbar n\textcolor{red}{-\frac{\hbar}{2}}$ \tabularnewline \hline
$L^\mathrm{(\pi)}$ & $\textcolor{blue}{\hbar n}=L^\mathrm{(K)}$ & $L^\mathrm{(K)}\textcolor{red}{-\frac{\hbar}{2}}$ & $L^\mathrm{(K)}\textcolor{red}{-\hbar}$ & $\textcolor{blue}{2\hbar n}=L^\mathrm{(K)}$ & $L^\mathrm{(K)}\textcolor{red}{-\hbar}$ \tabularnewline \hline
\end{tabular}
\caption{
\textit{Quantization concealment of angular momentum} --
Comparison of angular wavenumber, $m$, canonical angular momentum, $L$, and its derived kinetic angular momentum, $L^\mathrm{(\pi)}$, of the \textit{primary planar} eigen-modes\,\cite{planar_note}
 of a central force system, at the \textit{same} discrete orbital energy, $\hbar\omega_n$, under different dimension of quantization (DQs).
Notice an $\hbar$ ``concealment'' of $L$ in the 3D-quantized description that leads to underestimation of the kinetic angular momentum, $L^\mathrm{(\pi)}$, from its physical value, $L^\mathrm{(K)}=\hbar n$.
A similar $\hbar$ quantization concealment in $L^\mathrm{(\pi)}=L^\mathrm{(K)}-\hbar$ is also present in the magnetic field-driven cyclotron motion under 2D quantization, or with additional $z$-direction quantization ($+z$Q) in the cylindrical coordinate system.
}
\vspace{-0.4cm}
\label{tab:e_vs_m}
\end{table}

Indeed, Fig.\,\ref{fig:circular_2d_panels}(c) confirms that from $\langle L\rangle$ the (slightly over-estimated) angular velocities, $v_\phi^{(L)}\equiv\frac{\langle L\rangle}{(m_e r_0^2)}$, using a (slightly under-estimated) radius, $r_0=a_0 n_0^2$, is still \textit{slower} than the actual wave packets in 2D quantization, and even twice slower in 3D quantization.
Altogether, as summarized in Tab.\,\ref{tab:e_vs_m}, these analysis unambiguously reveals a generic shortcoming of wave mechanics: a \textit{systematic} $\hbar\frac{d-1}{2}$ concealment of kinetic $L^\mathrm{(K)}$ from canonical $L$ under $d$-dimensional quantization.

Mathematically, this concealment can be traced to the $\frac{d-1}{r}\partial_r$ contribution to the $d$-dimensional Laplacian in Eq.\,\ref{eq:3dradial} and \ref{eq:2dradial}, that has no direct classical interpretation and purely results from quantization beyond the dynamically active angular degree of freedom.
For \textit{bound} eigen-modes with $\alpha$ radial nodes, the resulting relation, $n=\frac{d-1}{2}+l\mathrm{(or}~|m|\mathrm{)}+\alpha$, of this term imposes a systematic $\frac{d-1}{2}$ reduction in quantizing angular wavenumbers for any \textit{given} $n$ or $\omega_n$.

As a matter of fact, such quantization concealment of kinetic $L^\mathrm{(K)}$ in canonical $L$ has long been recognized for the cyclotron motion under a magnetic field.
Under the symmetric gauge, the canonical angular momenta, $\hbar m$, of the primary eigen-mode $\ket{n,m=n}$ of the $n$-th Landau level\,\cite{LandauLifshitz1977} is systematically $\frac{\hbar}{2}$ short of the expected $\frac{L^\mathrm{(K)}}{2}$ value under 2D quantization\,\cite{Johnson1949,kitadono2020role,Van2020,WAKAMATSU2021168647}, equivalent to an $\hbar$ concealment of $L^\mathrm{(K)}$.
(3D quantization via the cylindrical coordinate system does not alter the $\frac{1}{r}\partial_r$ term and thus receives the same concealment.)

\textit{Gauge freedom of canonical momenta} --
This overall shift in angular wavenumbers (in each rotating direction) can be viewed as part of the \textit{generic} ``gauge freedom'' of \textit{canonical} momenta.
As a simple illustration, observe a planewave, $e^{i(\mathbf{k}\cdot\mathbf{x}-\omega t)}$, in a moving frame of relative velocity $\mathbf{u}$, such that $(t^\prime, \mathbf{x}^\prime)=(t,\mathbf{x}-\mathbf{u}t)$.
The observed planewave would appear as $e^{i(\mathbf{k}^\prime\cdot\mathbf{x}^\prime-\omega^\prime t^\prime)}$ with $(\omega^\prime,\mathbf{k}^\prime)=(\omega-\mathbf{k}\cdot\mathbf{u},\mathbf{k})$.
Clearly, without an additional gauge transformation, $\mathbf{k}^\prime \to \mathbf{k}^\prime - \frac{m_e}{\hbar}\mathbf{u}$, the \textit{canonical} momenta, $\mathbf{p}=-i\hbar\nabla$, would transform as $\mathbf{p}^\prime=\mathbf{p}$, in contrast to the \textit{kinetic} momenta, $\mathbf{p}^{\mathrm{(K)}\prime}=\mathbf{p}^\mathrm{(K)}-m_e\mathbf{u}$.
That is, just as in classical mechanics, canonical momenta only correspond to kinetic momenta under particular gauge.

Of course, such gauge freedom of canonical momenta should \textit{not} affect local wave packets' dynamical evolution.
Designed to mimic the classical trajectory of motion, a wave packet, $\int\psi(\mathbf{k}=\mathbf{k}_0+\Delta \mathbf{k})e^{i(\mathbf{k}\cdot\mathbf{x}-\omega t)}d^3k$, acquires physically relevant time evolution only though its ($\Delta \mathbf{k}$-dependent) \textit{relative} distribution, independent of the central wavenumber $\mathbf{k}_0$, or an overall gauge shift of $\mathbf{k}$ within $\psi$.
Indeed, this wave packet's group velocity, $\mathbf{v}^\prime=\nabla_{k^\prime}~\omega^\prime=\nabla_k(\omega-\mathbf{k}\cdot\mathbf{u})=\mathbf{v}-\mathbf{u}$, still corresponds to the gauge invariant kinetic momentum.

Analogously, as demonstrated above, a wave packet's orbiting motion and its kinetic $L^\mathrm{(K)}$ are gauge invariant.
However, since in quantum theories the gauge of the quantized canonical $L$ is already fixed to give the observed discrete energy levels, there is \textit{no more} freedom to maintain its \textit{direct} correspondence to the kinetic $L^\mathrm{(K)}$.
The above analysis shows that such gauge turns out to systematically conceal a $\hbar\frac{d-1}{2}$ piece of kinetic $L^\mathrm{(K)}$ from canonical $L$.

\begin{figure}[!t]
  \centering
  \includegraphics[width=\linewidth]{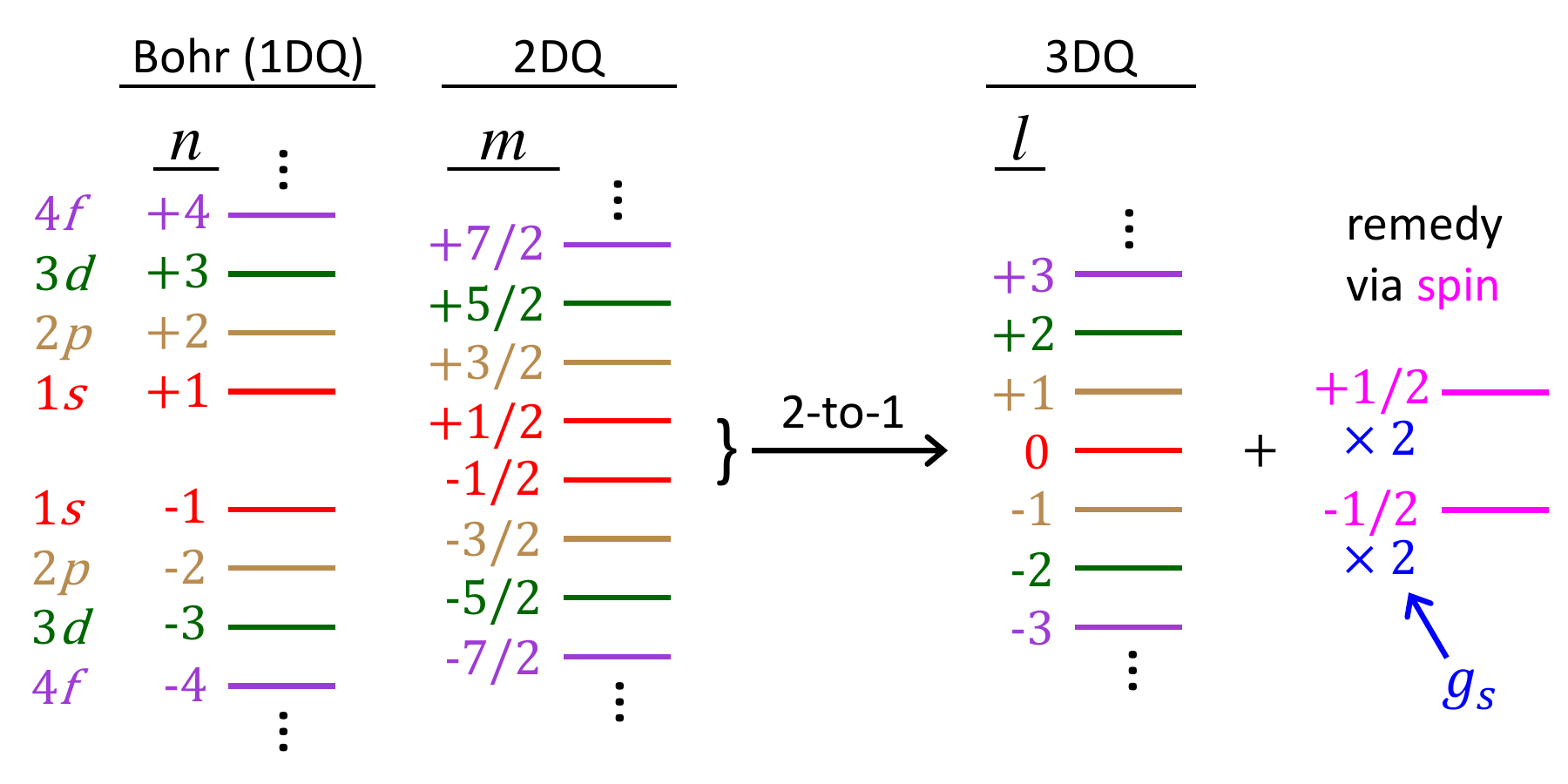}
  \vspace{-.7cm}
  \caption{\textit{Quantization concealment of angular momentum} --
  Comparison of primary planar eigen-modes' canonical angular wavenumber between Bohr, 2D-, and 3D-quantization given in Tab.\,\ref{tab:e_vs_m}.
  Notice the collapse of two eigen-modes of opposite directions into one in the $1s$ eigen-mode due to the angular momentum concealment, typically remedied by introducing a 1/2-spin with $g_s=2$.}
  \label{fig:L_mapping}
  \vspace{-0.4cm}
\end{figure}

\textit{``Zero'' angular momentum of the 1s orbital} --
This angular momentum concealment offers a very natural resolution to the counterintuitive ``zero'' angular momentum of the atomic $1s$ orbital in 3D quantized theories.
How can an electron maintain its position around 1 Bohr radius from the nucleus without an kinetic angular momentum?
Or equivalently, how to make sense the large \textit{kinetic} energy,
$\langle \frac{p^2}{2m_e} \rangle = -\omega_n$, with an identically \textit{zero} current density?
Table\,\ref{tab:e_vs_m} and the illustration in Fig.\,\ref{fig:L_mapping} shows that the physical \textit{kinetic} $L^\mathrm{(K)}$ of the $1s$ orbital is actually $\hbar$, in \textit{perfect consistency} with Bohr's model and its expected radius.
The standard \textit{canonical} $\langle L\rangle =\hbar l=0$ simply does \textit{not} reflect $L^\mathrm{(K)}$ due to the generic quantization concealment.

\textit{Incomplete minimal coupling} --
Given the above concealment of kinetic $L^\mathrm{(K)}$ from canonical $L$, a direct use of $L$ in place of $L^\mathrm{(K)}$ is obviously inaccurate, for example, in the standard ``minimal coupling'' to an external magnetic field, $\mathbf{B}$.
In such formulation, contributions, $\frac{1}{2m_e}(\mathbf{p}+e\mathbf{A})^2$, $\frac{e}{2m_e}\mathbf{L}\cdot\mathbf{B}$, and $e\mathbf{A}\cdot\mathbf{j}$, all explicitly use the gauge-shifted canonical momenta, $\mathbf{L}$, or its corresponding $\mathbf{p}$ or $\mathbf{j}$.
Therefore, the coupling to the magnetic field in Hamiltonian is inevitably \textit{underestimated} (the $1s$ eigen-modes being a good example).
In turn, the resulting eigen-frequencies are inadequate and require additional patch-up, such as spin-associated current.

\textit{Magnetic moment and anomalous Zeeman splitting} --
To correctly describe the physical current and magnetic moment, it is necessary to recover the physical kinetic $L^\mathrm{(K)}$ consistent with wave packets' motion.
Under 2D quantization $L^\mathrm{(K)}$ is simply a ``stretched'' $L$ (c.f. left two columns of Fig.~\ref{fig:L_mapping}),
\begin{equation}
L^\mathrm{(K)}=\sum_{m\in \mathbb{Z}+\frac{1}{2}} \ket{m}\left[\hbar m\left(1+\frac{1}{2|m|}\right)\right]\bra{m}
\label{eq:fix_2D}
\end{equation}
In turn, the associated physical current,
\begin{equation}
\mathbf{j}^\mathrm{(K)}=\frac{1}{2}\hat{\phi}\times\bigg(L^\mathrm{(K)}\frac{\mathbf{r}}{r^2}+\frac{\mathbf{r}}{r^2}L^\mathrm{(K)}\bigg),
\label{eq:fix_current_2D}
\end{equation}
produces a larger magnetic moment, $\frac{e}{2m_e}L^\mathrm{(K)}$, that correctly reproduces the observed anomalous Zeeman splitting, $\pm\frac{e}{2m_e}L^\mathrm{(K)}B$, of the planar modes\,\cite{zeeman1897vii,michelson1889feasibility,pauli1924frage,rabi1938new} even for the $1s$-orbitals.
That is, instead of the gauge-shifted canonical $L$, the planar orbiting wave packets' time evolution already gives a kinetic current whose \textit{orbital} magnetic moment matches their observed anomalous Zeeman splitting, without the need for phenomenological introduction of spins.

\textit{Spin as an excellent remedy} --
Nonetheless, from the above consideration of kinetic \textit{orbital} angular momentum, $L^\mathrm{(K)}$, one sees that the phenomenological introduction of spins is an extremely effective remedy to recover the proper physical properties of $L^\mathrm{(K)}$ for bound eigen-modes.
Indeed, the additional spin-associated current density $\mathbf{j}_s$ in Eq.\,\ref{eq:NR_current}, while differs in detail, very nicely captures the overall structure of $\mathbf{j}^\mathrm{(K)}$ in Eq.\,\ref{eq:fix_current_2D}.
Not only it contains an unusual \textit{non-commuting} $x$- and $y$-components as in $\mathbf{j}^\mathrm{(K)}$, but it also produces the unaccounted $\frac{e}{2m_e}\hbar$ piece of the magnetic moment in standard 3D quantized theories.
Particularly, even with spin of arbitrary size, $S$, in linearization, Belinfante's conjecture~\cite{belinfante1953}, $g_s=S^{-1}$, guarantees a spin magnetic moment, $\frac{e\hbar}{2m_e}$, from $\mathbf{j}_s$, in agreement with concealed piece of the magnetic moment.

The employment of the extended ``spinor'' Hilbert space\,\cite{Wilkes_2020,Hurley1971} further illustrates the effectiveness of spins as a remedy to the quantization concealment, when attempting to recover kinetic $L^\mathrm{(K)}$ under 3D quantization.
As illustrated in Fig.\,\ref{fig:L_mapping}, the concealment in 3D is too severe that it collapses the two $s$-orbitals' planar modes (with opposite orbiting directions) into one.
That is, eigen-modes in 3D quantization become \textit{generically} inadequate to represent the kinetic angular momentum.
The spinor extension nicely recovers from the over-concealment-induced collapse of Hilbert space in standard quantum theories.

Furthermore, the spinor extension also offers a simple solution to the nasty issue of complete Hilbert space for 3D rotations.
To illustrate this, let's expand the 2D planar modes, $\ket{v_{n|m|}}\ket{m}|_{m=l+\frac{1}{2}}$, to their 3D counterparts\,\cite{Hunter1999,Pav_i__2007}, \begin{equation}
\bra{ r,\theta,\phi}\big(\ket{u_{nl}}\ket{j,j_z=j}\big)\big|_{j=|m|} \propto u_{nl}(r)\sin^{|m|}(\theta)e^{i\phi |m|},
\label{eq:planar_mode_3D}
\end{equation}
in an attempt to recover from the over-concealment collapse of modes under 3D quantization.
In principle, the complete ($2j+1$)-dimensional Hilbert space for each $j$ can be generated through declination, $D_\mathbf{n}$, of $\ket{j,j}$ along all possible directions  $\mathbf{n}$ in the solid angle, $\Omega$.
(See Ref.\,\cite{SM} for demonstrations for an integer $j$.)
However, for half-integer $j$ one finds it challenging to form an adequate complete Hilbert space with the expected ($2j+1$) dimension of linear independence.
This is because the the twisted boundary cut (TBC) between $\phi=2\pi$ an 0 in Eq.\,\ref{eq:planar_mode_3D} moves its location under 3D rotations of $\ket{j,j}$.
Thus, unless an unusual Hilbert space that relieves such geometrical limitation (for example via ``symmetrizing'' all TBCs over all possible 3D rotation) can be designed, the presence of TBCs in the generated Hilbert space generally breaks the completeness requirement of the rotational group.

The employment of spinors very naturally resolves this difficulty.
On the expense of losing some spatial information associated with spinors, the standard construction of $\ket{j=l+\frac{1}{2},j_z}$ by adding a spin-1/2 spinor $\ket{\frac{1}{2},\pm\frac{1}{2}}$ to $\ket{l,m}$ with integer $l$ very effectively retains the ($2j+1$)-dimensional Hilbert space for each $j$, as revealed by the split levels under a magnetic field.
Again, one sees that introduction of spins and their spinor Hilbert spaces offers an excellent remedy to the quantization concealment of angular momentum in facilitating a simple description with observed group properties.

\textit{Resolution to current puzzles of spins} --
From this perspective of spins being an excellent remedy, the angular momentum concealment must then offer a natural resolution to the current puzzles about spins.
Indeed, as clarified above, the spin-associated orbiting current from Eq.\,\ref{eq:NR_current} nicely recovers the overall effect of the missing piece of $\mathbf{j}^\mathrm{(K)}$.
Therefore, for the \textit{bound eigen-modes}, the dynamical origin of the mysterious spin-associated current is nothing but the central force (or the Lorentz force for the cyclotrons) that gives rise to the kinetic angular momentum, $L^\mathrm{(K)}$.

Furthermore, while the physical currents from the kinetic $L^\mathrm{(K)}$ and the gauge-shifted canonical $L$ are inertial frame dependent, the quantization concealed \textit{fixed} $\frac{d-1}{2}$ \textit{difference} between $L^\mathrm{(K)}$ and $L$ is frame invariant.
Explicitly, since the frame transformed equations of motion still have the same dimension of quantization, their bound eigen-modes would suffer the same amount of angular momentum concealment.
This \textit{effortlessly }explains the puzzling frame  invariance of the spin angular momentum and its associated spatial current.

\textit{Absence of spin magnetic moment in unbound particles} --
Note that above considerations on angular momentum concealment are all for \textit{bound} eigen-modes.
For unbound eigen-modes, the concealment is negligibly small far from the center of the force.
Thus, introduction of spin-associated current to unbound modes is unnecessary and risks producing \textit{unphysical} effects.
For example, inside the wave packet of a nearly freely propagating wave packet, there is absolutely \textit{no} dynamical drive for a \textit{spatially  circulating} current.
Thus, $\mathbf{j}_s$ resulting from linearized equations of motion should be taken as `information' for their mathematical integrity, but \textit{not} as a physical current capable of generating magnetic moments.

\textit{Predictions} --
One thus arrives at a seemingly surprising conclusion that unbound particles actually do not carry spin-magnetic moments.
Currently, all measurements of electronic spin magnetic moments involve trapping electrons into an orbiting motion either in atoms\,\cite{Rivic2023,Phipps1927} or under a magnetic field\,\cite{Elzerman2004,Fan2023,gbkc-6nlj}.
The system thus inevitably contains orbital magnetic moment experimentally inseparable from the spin contribution.
The above analysis predicts that \textit{direct} measurement of spin magnetic moments, for example through Stern-Gerlach experiment\,\cite{Gerlach1922} on electrons, would \textit{not} observe spin-magnetic moment-induced splitting (or any other effects) along the direction of the field gradient.

The finite kinetic angular momentum of $1s$ orbital also predicts that under magnetic fields its eigen-modes should have zero probability along the $z$ axis defined by the external field, in great contrast to the even angular probability in the standard theories.
(The same conclusion can be drawn from the $\mathbf{j}_\mathrm{s}$-associated dynamics in Eq.\,\ref{eq:NR_current}.)
This should be verifiable through accurate spectral intensity analysis that reflects the transition matrix elements, an interesting direction of future theoretical and experimental investigation.

\textit{Discussions} --
Nonetheless, even in the absence of spin-magnetic moment for unbound particles, keeping track of spins as an ``internal property'' still seems valuable.
This is because currently emission, absorption, or detection of free particles still involves dynamical processes of trapping them, and thus keeping track of the concealed piece of kinetic $\mathbf{L}^\mathrm{(K)}$ is necessary as part of the conserved $\mathbf{L}^\mathrm{(K)}$ of the entire system.
Attaching such information to particles during their free propagation (via an extended spinor Hilbert space) appears to be the most natural option.
(One just have to resist the temptation of interpreting it as physical current or magnetic moment.)

Throughout this study, our consideration focuses only on the microscopic \textit{dynamical} origin of spin-related phenomena.
Spin-related \textit{statistical} aspects, such as number of quantized modes between the planar modes, their allowed electronic occupation number, or electrons' anti-symmetric indistinguishability, are beyond the scope of this study.
Some in-depth considerations on these issues are available in Ref.\,\cite{Hansson2001,Haldane1991,Pauli1940,Feldmeier2000,XiaoDi2005,Subramanyan2025}.

Still, the equivalent description in 2D quantization \textit{does} clearly demonstrate that concerning physical dynamics, quantization is only necessary for the dynamically active dimension, while quantizing other dimensions only facilitates convenient statistics.
This in turn supports the notion\,\cite{Sommerfeld1916,Wilson1915} that a fixed-scaled uncertainty, such as $\hbar$, may originates from 
conserved quantities emerging from
underlying rapid dynamics.

Finally, for clarity our illustration only makes use of non-relativistic systems.
Nonetheless, since the revealed concealment is only associated with the dimension of quantization, details of dynamics, such as relativistic effects, should not affect it.
Indeed, identical quantization concealment can be shown\,\cite{SM} via Dirac equation as well.
Careful consideration for effects of relativistic \textit{dynamics} on level quantization would be an interesting topic for future studies.

\textit{Conclusion} --
We expose a long-overlooked generic shortcoming of wave mechanics -- `quantization concealment of angular momentum' of its \textit{bound} eigen-modes that renders incomplete the minimal coupling to magnetic fields.
Recovering the physical kinetic \textit{orbital} angular momentum resolves the counterintuitive ``zero'' angular momentum of the 1$s$ orbital and reproduces the observed anomalous Zeeman splitting.
Viewing spins as remedy to the concealment offers natural explanation for 1) the dynamical origin of the spin magnetic moment and 2) frame invariance of spin angular momentum.
It also implies \textit{absence} of spin magnetic moment for free particles, a strong prediction directly verifiable through Stern-Gerlach experiments on electrons.

\begin{acknowledgments}
The authors acknowledge helpful discussions with Matheus de Sousa and Jie Wang.
This work is supported by National Natural Science Foundation of China (NSFC) \#12274287 and \#12042507 and Innovation Program for Quantum Science and Technology 2021ZD0301900.
\end{acknowledgments}

\bibliography{references}
\end{document}


\begin{CJK*}{UTF8}{gbsn}

\title{Supplemental Material for ``Quantization concealment of angular momentum: dynamic origin of spins"}

\author{Wenjie Long (\CJKfamily{gbsn}龙文婕)}
\affiliation{School of Physics and Astronomy, Shanghai Jiao Tong University, Shanghai 200240, China}
\author{Wei Ku ({\CJKfamily{bsmi}顧威})}
\altaffiliation{corresponding email: weiku@sjtu.edu.cn}
\affiliation{School of Physics and Astronomy, Shanghai Jiao Tong University, Shanghai 200240, China}
\affiliation{Key Laboratory of Artificial Structures and Quantum Control (Ministry of Education), Shanghai 200240, China}
\affiliation{Shanghai Branch, Hefei National Laboratory, Shanghai 201315, China}

\date{\today}
\maketitle
\end{CJK*}

\setcounter{equation}{0}
\renewcommand{\theequation}{S\arabic{equation}}
\setcounter{figure}{0}
\renewcommand{\thefigure}{S\arabic{figure}}
\setcounter{table}{0}
\renewcommand{\thetable}{S\arabic{table}}

\section{Spin-induced spatial current in relativistic systems}

Similar to the non-relativistic systems, for a relativistic system described by the Dirac Hamiltonian,
\begin{equation}
H=\tau_1 \sigma \cdot \mathbf{p}+ \tau_3 m_e +V(r),
\label{eq:H_Dirac}
\end{equation}
denoted via Pauli matrices~\cite{Pauli1927} $\sigma$ and $\tau$ belonging to separate Hilbert spaces,
the spatial current contains a spin-induced contribution~\cite{Gordon1928},
\begin{equation}
\mathbf{j}=\frac{\hbar}{2m_e}\nabla\times(\psi^\dagger\tau_3\sigma\psi),
\label{eq:R_current}
\end{equation}
as well.
Under external magnetic field, $\mathbf{B}=\nabla\times \mathbf{A}$, this gives an extra energy,
\begin{equation}
U_s =-\int(-e)\mathbf{j}\cdot \mathbf{A}~d^3r
=-\mathbf{B}\cdot\int\psi^\dagger(\frac{-e\hbar}{2m_e}\tau_3\sigma)\psi~d^3r,
\label{eq:NR_moment}
\end{equation}
equivalent to a magnetic momentum of size $\frac{e\hbar}{2m_e}$ or a gyromagnetic ratio 2 for a 1/2-spin.

\section{Quantization concealment in relativistic systems}

In correspondence to primary modes in non-relativistic case, consider the Dirac eigenfunction in the $\kappa=-(j+\frac{1}{2})$ sector, where the large component has orbital angular momentum and spin aligned,
\begin{equation}
\Psi_n(\mathbf r)=
\begin{pmatrix}
u_{n}(r)\,y^{j-\frac{1}{2}}_{jm_j} \\ iv_{n}(r)y^{j+\frac{1}{2}}_{jm_j}.
\end{pmatrix}
\end{equation}
Here $u_n(r)$ and $v_n(r)$ are the radial functions of the upper and lower components, respectively, and $y^l_{jm}$ are the \textit{spinor spherical harmonics} with orbital angular momentum $l$, total angular momentum $j$, and its projection $m$. 

Especially, for $m_j =j$
\begin{equation}
y^{j-\frac{1}{2}}_{jj}=
\begin{pmatrix}
Y_{j-\frac{1}{2}}^{j-\frac{1}{2}}\\0
\end{pmatrix},\quad
y^{j+\frac{1}{2}}_{jj}=\frac{1}{\sqrt{2j+2}}
\begin{pmatrix}
-Y_{j+\frac{1}{2}}^{j-\frac{1}{2}}\\\sqrt{2j+1} Y_{j+\frac{1}{2}}^{j+\frac{1}{2}}
\end{pmatrix},
\end{equation}
where $Y_l^m$ are \textit{spherical harmonics}.

Denoting $l=j-\frac{1}{2}$, Eq~\ref{eq:H_Dirac} is now reduced to a radial equation eigen-equation~\cite{thaller2013dirac},
\begin{equation}
\begin{pmatrix}
m_e + 
V(r) &  -(\partial_r + \frac{l+2}{r})\\
 \partial_r - \frac{l}{r} & -m_e + 
V(r)
\end{pmatrix}
\begin{pmatrix}
u_{n}(r) \\ v_{n}(r)
\end{pmatrix}
=\omega_{n}
\begin{pmatrix}
u_{n}(r) \\ v_{n}(r)
\end{pmatrix}
\label{eq:radial_Dirac_3D}
\end{equation}

In 2D, quantize Dirac Hamitonian,
\begin{equation}
\mathbf{\sigma} \cdot \mathbf{p}=
\begin{pmatrix}
0 & -ie^{-i\theta}(\partial_r - \frac{i}{r}\partial_\theta) \\-ie^{i\theta}(\partial_r + \frac{i}{r}\partial_\theta) & 0
\end{pmatrix}
\end{equation}

The eigenfunction,
\begin{equation}
\Psi_n(\mathbf r)=
\begin{pmatrix}
g_{n}(r)e^{im\theta}\\0\\0 \\ -if_{n}(r)e^{i(m+1)\theta}.
\end{pmatrix}
\end{equation}
\begin{equation}
\begin{pmatrix}
m_e + 
V(r) &  -(\partial_r + \frac{m+1}{r})\\
\partial_r - \frac{m}{r} & -m_e + 
V(r)
\end{pmatrix}
\begin{pmatrix}
g_{n}(r) \\ f_{n}(r)
\end{pmatrix}
=\omega_{nm}
\begin{pmatrix}
g_{n}(r) \\ f_{n}(r)
\end{pmatrix}.
\label{eq:radial_Dirac_2D}
\end{equation}
produces Eq.~\ref{eq:radial_Dirac_3D} upon setting $m=l+\frac{1}{2}$, $g_{n}(r)=r^\frac{1}{2}u_{n}(r)$ and $f_{n}(r)=r^\frac{1}{2}v_{n}(r)$.

As expected, quantization concealment of the angular momentum occurs as well.

\section{Non-planar eigen-modes, $\ket{l,m}$, as rotating planar mode, $\ket{l,l}$, in declination}

\begin{figure}[h]
  \centering
  \includegraphics[width=1\linewidth]{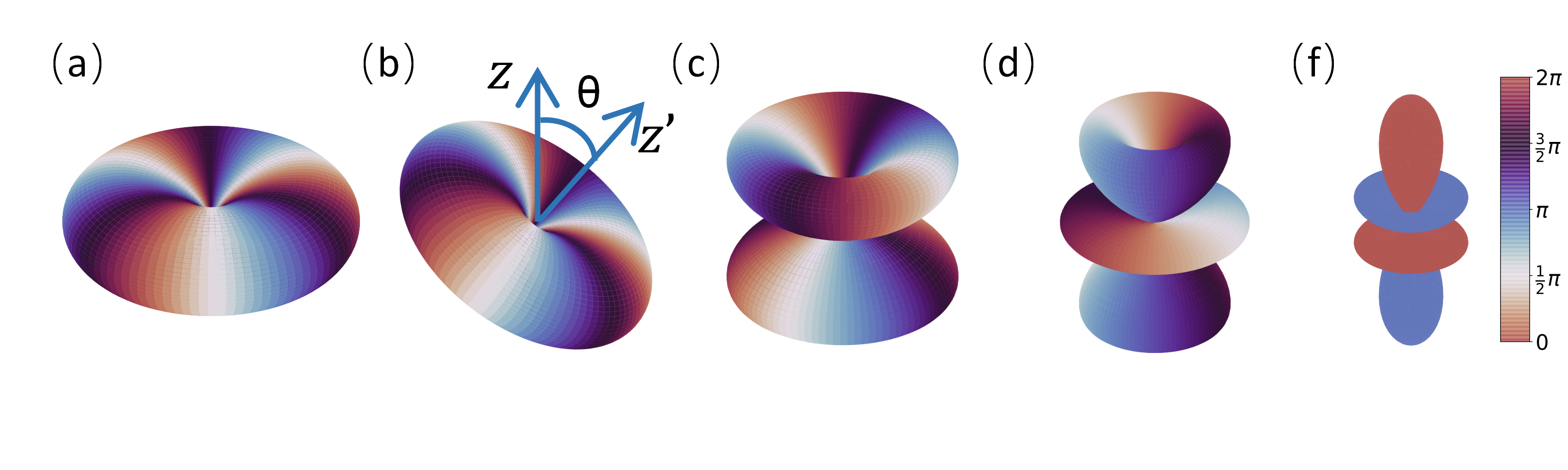}
  \vspace{-.7cm}
  \caption{\textit{Non-planar eigen-modes, $\ket{3,m<3}$, from precessing planar modes, $\ket{3,3}$, under declination} -- (a) Planar eigen-modes $\ket{3,3}$, (b) the same under declination of angle $\theta^{(d)}_m$, and its precession represented by superposition with a relative phase, $e^{im\phi}$, for (d) $m=2$, (e) $m=1$, and (f) $m=0$.}
  \label{fig:3drotation}
  \vspace{-0.4cm}
\end{figure}

Here we demonstrate generation of non-planar modes, $\ket{l,m}|_{|m|<l}$, from a precessing planar eigen-mode, $\ket{l,l}$, in declination.
Consider, for example, $l=3$, the planar eigen-mode,
\begin{equation}
    Y_l^l \propto \sin^l(\theta)e^{il\phi},
\end{equation}
in a plane under a declination angle, $\theta^{(d)}_m=\arccos{(\frac{m}{l})}$, shown in Fig.\,\ref{fig:3drotation}.
Its precession over a long time scale is then represented by a superposition with a coherent relative phase $e^{im\phi}$.
One sees that the resulting modes are exactly the non-planar eigen-modes.

\bibliography{references}